\documentclass[preprint]{aastex}
\usepackage{xspace}
\usepackage[onecolumn]{emulateapj5}
\def\ergg{${\rm \hbox{erg}\ g^{-1}}$\xspace}

\def\sun {${}_\odot$}
\def\Msun{M\sun\xspace}

\def\mch{${\rm M_{ch}}$\xspace}
\def\Mch{\mch}
\def\nifsx{${}^{56}$Ni\xspace}
\def\cofsx{${}^{56}$Co\xspace}
\def\fefsx{${}^{56}$Fe\xspace}
\def\cf{{\it cf.}\xspace}

\def\ie{{\it i.e.}\xspace}

\shortauthors{Pinto \& Eastman}
\shorttitle{SN~Ia Width-Luminosity Relation}

\begin{document}

\title{The Type Ia Supernova Width-Luminosity Relation}

\author{Philip A. Pinto\altaffilmark{1}}
\affil{Steward Observatory, The University of Arizona, Tucson, AZ 85721 USA}
\email{ppinto@as.arizona.edu}
\and
\author{Ronald G. Eastman\altaffilmark{2}}
\affil{Lawrence Livermore National Laboratory, Livermore, CA 94550 USA}
\email{reastman@llnl.gov}
\altaffiltext{1}{Lawrence Livermore National
Laboratory, Livermore CA 94550 USA}
\altaffiltext{2}{Department of Astronomy and Astrophysics,
University of California, Santa Cruz, Santa Cruz CA 95064 USA}

\begin{abstract}
Observations of type Ia supernov\ae\ at high redshifts have become an
important tool for studying the geometry of the universe. The relation
between the duration of the peak phase of a type Ia supernova's lightcurve
and its luminosity (broader is brighter) forms the cornerstone of this
measurement, yet it is a purely empirical relation. In this paper we show
that the relation is a natural consequence of the radiation transport in
type Ia supernovae and suggest constraints on the nature of the explosions
which arise from our interpretation of the observed relation. The principle
parameter underlying the relation is the mass of radioactive \nifsx
produced in the explosion. The relation is shown to be relatively
insensitive to most other parameters in Chandrasekhar-mass explosions.
Cosmological results are thus unlikely to suffer from systematic effects
stemming from evolution in the explosions' progenitors.
\end{abstract}


\section{Introduction}
Because they are among the brightest optical explosions in the universe,
the promise of employing type Ia supernov\ae\ (SNe~Ia) as cosmological
probes has long been recognized \citep{Kowal68}. In the past few years,
this promise has been realized by two groups
\citep{Schmidtetal98,Perlmutteretal99a} who have used observations of
SNe~Ia to determine luminosity distances to supernov\ae\ at redshifts
beyond $z=1$. The results have been surprising: ours is a universe which is
expanding at an accelerating rate. In several ways this result is in accord
with our expectations. The observations imply that the age of the universe
is near 14~Gy, comfortably older than the age of its contents. They also
imply that the density of ordinary matter is well below the critical
density, near $\Omega_{matter}=0.3$, and this too is in agreement with most
dynamical measurements \cite{PerlmutterTW99}. Perhaps the most exciting
result, and certainly the most unnerving and controversial, is that the
geometry of the universe is today dominated by an energy density which does
not arise from ordinary matter but which is best fit by a vacuum energy
density for which there are no satisfying explanations
\citep{Riessetal98,Garnavichetal98}.

To be really useful in the classic cosmological test of the
redshift/luminosity-distance relation, the luminosity of a standard candle
must have a small dispersion. The set of ``typical'' SNe~Ia used for
cosmology exhibits a range of almost a factor of three in peak
brightness. This would not be sufficient to enable the current sample,
which extends to a redshift of $z\sim1$, to place useful limits on
cosmological acceleration; the observed departure from a flat,
matter-dominated universe amounts to three tenths of a magnitude
($\sim30$\% in brightness). SNe~Ia are useful as cosmological probes only
because what variations are observed can, for the most part, be calibrated
out, leading to a usefully small dispersion in corrected brightness of
about 0.15 mag.

\citet{Phillips93} demonstrated from a set of nearby SNe~Ia that the rate
of decline from peak in the optical lightcurve is correlated with peak
brightness. Using the larger and more uniform data set provided by the
Cal\'an-Tololo supernova survey, \citet{Hamuyetal96b} showed that this
correlation exists in the B-, V-, and I-band lightcurves (though,
interestingly, with different slopes than those determined by
\citealt{Phillips93} from the nearest supernovae). They found that not only
the decline rate but also the overall width of the lightcurve centered
on maximum light is
correlated with peak luminosity.  Using this same sample, \citet{RiessPK95}
also showed that the lightcurve shape could be used as a predictor of
luminosity. Perhaps most suggestive from a physical point of view, using a
sample of high-redshift supernov\ae\ (for which very early time
observations are more easily obtained), \citet{Perlmutteretal97} have shown
that not only is there a correlation of the lightcurve shape with
luminosity but that, near peak, the lightcurve is the {\em same} in all
supernov\ae\ up to a correlated scaling in time and luminosity, although
quantitative errors for this relation have yet to be published.

While this width-luminosity relation (henceforward WLR) is very well
substantiated observationally, it is an {\em empirical} relation which has
remained largely unexplained. \citet{Hoflichetal96} showed that many
current explosion models come close to this relation and suggested that the
relation was determined primarily by the precise nature of the explosion
itself.  In this paper we provide the rudiments of a physical explanation
for the WLR which is based upon the physics of radiation flow in SNe~Ia,
one in which the details of the explosion itself are largely masked. If
true, this argues strongly against any significant bias arising in the WLR
due to evolutionary effects in the supernova progenitors.

SNe~Ia are widely believed to result from the thermonuclear incineration of
an accreting carbon/oxygen white dwarf.  Beyond this, however, little is
known with any degree of precision. Present uncertainties include the
nature and evolution of the progenitor system, the mass of the dwarf at
ignition, and the physics of the subsequent nuclear burning. Given this
state of our knowledge, it would at first blush appear hopeless to explain
the WLR from physical principles. It may be, however, that the calibrating
relation arises naturally from conditions present in many or all such
explosions of accreting dwarfs and that the details of progenitor evolution
and burning physics are of little consequence to the WLR, though
fascinating in their own right.

For want of any more certain starting point, we adopt this point of view
and examine how far it may take us toward an understanding of the WLR for
SNe~Ia. In the first section, we reprise some of the physics of radiation
transport in SNe~Ia explored in two previous papers
(\citealp{PintoE00a,PintoE00b}, hereafter PE~I \& II). We then develop a
schematic description of the lightcurve physics which illustrates the two
main properties of the explosion which determine the WLR: the mass of
radioactive \nifsx produced in the explosion and the rate of change of
$\gamma$-ray escape. In the following sections we then present a sample set
of lightcurve calculations which demonstrate the essential correctness of
this view, and demonstrate that the observed WLR can be reproduced with
this physics. We conclude with speculations on the probability that
evolutionary effects can materially alter the relation observed in the
local universe and lead to false cosmological conclusions.

\section{Theory \label{theorysec}}

The WLR is the observation that the greater the luminosity at maximum
light, the greater the width of the light curve, or, equivalently, the
slower the rate of decline from maximum. As discussed in PE~I, the
lightcurve of a SN~Ia is determined by a competition between three effects:
the deposition of energy from radioactive decay, the adiabatic
conversion of internal energy to kinetic energy of expansion, and the
escape of internal energy as the observed lightcurve.

Because the initial configuration is compact and at high density, the
explosion is at first tremendously optically thick. The timescale
associated with escape of internal energy is very long, and virtually all
of the energy liberated in the explosion is converted to kinetic energy on
a timescale approximately equal to the elapsed time.  Were it not for the
continuous energy input from decay of \nifsx and \cofsx produced in the
explosion, SNe~Ia would be both faint and short-lived. Fortunately, \nifsx
decay yields an energy of $3.0\times10^{16}\ \hbox{erg}\ \hbox{g}^{-1}$ and
\cofsx decay yields a further $6.4\times10^{16}\ \hbox{erg}\ \hbox{g}^{-1}$.

As the star expands, the optical depth decreases and the timescale over
which this energy can escape declines. While trapped, radiant energy is
constantly being converted into kinetic energy; the longer the radiation
remains in the supernova, the less energy will escape to power the
lightcurve. With a constant energy source, the luminosity would increase
monotonically, asymptotically to balance the deposition at zero optical
depth. In a real supernova, however, the deposition declines as radioactive
material decays away and as the $\gamma-$rays, in which the decay energy is
emitted, escape. Peak luminosity thus occurs roughly when the product of
the rising escape fraction and the decreasing deposition reaches a maximum.
This is complicated somewhat by the fact that the balance is not
instantaneous; the energy emerging as UV to IR photons at any time near peak
light has been deposited over a significant fraction of the elapsed time.
\citet{Arnett82} first demonstrated that at maximum light the instantaneous
bolometric luminosity is approximately equal to the instantaneous rate of
energy deposition by radiative decay.  Earlier, the expansion timescale is
short, the escape time is long, and the deposition rate is high.  More
internal energy goes into expansion than escapes, and a store of trapped
energy is built up. Subsequent to maximum light, more energy escapes than
is converted to kinetic energy, and the surviving store of trapped energy
is released faster than the deposition.

The WLR is thus seen to be a relation between the maximum-light luminosity
and the radiation {\em escape time}; the brighter the supernova, the longer
its escape time. Radiation escapes from an optically thick medium by
diffusion, and we can tentatively identify the escape time with the
diffusion time -- the average time it takes a photon to random walk its way
to the surface, given by the number of scatterings times the time of flight
between scatterings.

Following the usual random walk argument, the diffusion time can be
written as
\begin{equation}
t_d={\alpha R \tau_{eff}\over c}
\end{equation}
where $R$ is a characteristic size of the system (in this case, the radius
of the ejecta), $\tau_{eff}$ is the effective optical depth, and $\alpha$
is a constant of order unity. By the argument above, if the expansion
(elapsed) time and escape (diffusion) times are about equal at peak, we can
write
\begin{equation}
t_d=t={\alpha R \tau_{eff}\over c}={\alpha V t \tau_{eff}\over c}
\end{equation}
which gives a diffusion optical depth at peak of
\begin{equation}
\tau_{eff} \approx 30 \left({10^9\ \hbox{cm}\ {\rm s}^{-1} \over V}\right)
\end{equation}
(taking $\alpha\approx1$). This value is far smaller than the monochromatic
optical depth to most trapped photons, which is typically nearer to 3000 for
peak conditions in a SN~Ia. The resolution to this seeming paradox, of
course, is that the {\em typical} photons in an atmosphere do not carry the
flux -- radiation escapes where it can, at energies where the monochromatic
optical depth is smaller.

At stellar densities, there is tight collisional coupling between the
radiation field and the plasma, keeping the local photon distribution
function at its blackbody limit. While the bulk of the energy density
exists as UV photons, these photons are trapped by the high UV opacity. The
bulk of the flux is transported at longer wavelengths on the Rayleigh-Jeans
tail of the distribution, where the optical depth is typically less,
``side-stepping'' the higher UV opacity.  The lower density of these
photons is more than overcome by the decreased diffusion time they suffer,
and the appropriate opacity to use in describing the flux is the familiar
Rosseland mean.

For this picture to work, the relaxation rate of the trapped photon
distribution toward a blackbody must be short compared with the rate at
which energy escapes from the long-wavelength tail of the distribution.
This relaxation is usually mediated by collisions, and its rate must
therefore scale as the square of the density.  If the optical depth to the
majority of photons remains large (so that radiation remains trapped) while
the density decreases, eventually collisions will not be able to maintain
sufficient energy density in the long-wavelength tail of the photon
distribution where the optical depth is less; the system will depart from
local thermodynamic equilibrium (LTE).

PE~II demonstrated that this has indeed become the case in SNe~Ia by
maximum light.  The escape time is still mediated by the rate for
converting energy from short-wavelength, UV photons into long wavelength,
OIR photons. The difference from the more typical LTE situation arises from
the nature of opacity in rapidly-expanding media and the character of
atomic physics in iron-group ions.

In the homologous expansion which develops soon after explosion, the
velocity gradient is isotropic. Much as in the expanding universe, there is
a one-on-one mapping between path-length and redshift, and a photon
traversing the ejecta will accumulate a red shift in the local gas rest
frame which will bring it into resonance with line transitions.  One can
show \citep{EastmanP93} that under the conditions in SNe~Ia, the time a
photon spends scattering in a line is very short compared with the time it
spends moving between lines. For the purpose of estimating the escape time,
then, each line traversed thus acts as a single scattering, and the
distance measured in units of a photon mean free path (the optical depth)
will be just the number of optically thick lines traversed in accumulating
a given Doppler shift; the corresponding effective opacity is known as an
``expansion opacity'' (\cf
\citealp{EastmanP93,WagonerPV91,KarpLCS77,CastorAK75}).

Using simple analytic estimates, PE~I showed that the central temperature
in a \Mch explosion at $t<20$~days is $T\gtrsim13,000$~K and the peak of
the blackbody spectrum is in the UV near $\lambda\lesssim2200$~\AA. For an
iron-peak composition in the relevant range of temperature and density, the
spectral density of optically thick lines in the UV is very large and is a
strongly decreasing function of wavelength from the UV through the optical
(\cf Figure 2 of PE~II). As long as the Sobolev optical depth in these
lines remains large, the diffusion optical depth will not change as the
density decreases, allowing a departure from LTE not seen at depth in
static atmospheres where the optical depth is linearly dependent upon
density.  The large spectral density of lines at UV wavelengths is far too
great for there to be significant transport at UV energies; the time it
takes a UV photon to random walk to the surface is long compared to the
expansion time.  This is substantiated by the observation of very little UV
flux in maximum light SNe~Ia in spite of their high temperatures.

If the Rayleigh-Jeans tail of the photon distribution (at wavelengths where
significant transport can take place) could be re-populated by a mechanism
more efficient at low densities than collisional excitation, significant
flux could still be developed, leading to an escape time characterized by
the required effective optical depth near 30.

When a photon is absorbed in a line, the possible outcomes of this
interaction include: 1) re-emission in the same transition, followed
possibly by re-absorptions and emissions -- a cycle repeated many times
until escape from the red wing of the line; 2) collisional destruction,
transferring the photon energy by Coulomb interactions to a free electron;
and 3) fluorescence, where the excited state populated by absorption
de-populates by radiative decay to a state different from the original.  In
general, the electron density is too low in maximum light SNe~Ia for
collisional destruction to have much effect.  Thus, a UV photon absorbed on
the high energy (blue) side of a line will either be re-emitted on the red
side, or fluorescence will occur, with the photon replaced by several
longer-wavelength photons (usually), but occasionally by a shorter
wavelength photon (less often).  The efficiency for this conversion process
obviously depends on the number and strength of decay channels which would
result in the emission of an OIR photon. For the UV radiation trapped in
SNe~Ia, fluorescence is the most likely outcome. Fluorescence is much more
important in SNe~Ia than in other astrophysical objects because they are
composed in large part of iron-group elements with enormously complex
atomic structures and very high spectral densities of line transitions in
which fluorescence can occur.

A key point is this: the number of OIR transitions is greatest in the
neutral and low ionization states, and decreases with increasing
ionization. Thus the UV$\rightarrow$OIR fluorescence process is more
efficient in Co~II than in Co~III, and more efficient in Co~III than in
Co~IV.  Consequently, the conversion process is sensitive to the
ionization. In the approximate calculations reported below, we assume LTE
ionization and excitation despite densities which are too low for this to
be really true.  We shall thus use temperature as a proxy for ionization,
but the reader should keep in mind that this is an approximation, and it is
actually ionization which is the controlling variable.

In light of the importance of fluorescence, we can sketch the escape of
radiation from SNe~Ia in the following fashion. Imagine a store of
deposited energy, trapped in the form of UV photons. These UV photons are
created by the absorption and thermalization of decay $\gamma$-rays and
removed by conversion to kinetic energy and to OIR photons. For the energy
density of UV photons, $E_{UV}$, we can write the following schematic rate
equation:
\begin{equation}
{\partial E_{UV}\over\partial t} = \rho\dot{S}_{dep}(t) -
{E_{UV}\over t_e} - r_f E_{UV}
\label{UVRate}
\end{equation}
where $\rho\dot{S}_{dep}(t)$ is the time-dependent deposition rate,
$E_{UV}/t_e$ is the rate of conversion of radiation to kinetic energy, and
$r_f E_{UV}$ is the rate for converting UV to optical and infrared
energy. We have assumed here that the diffusion time suffered by the UV
energy is infinitely long; transport in the UV is unimportant. Similarly,
for the OIR photons one can write
\begin{equation}
{\partial E_{OIR}\over\partial t} = -{E_{OIR}\over t_e}
-{E_{OIR}\over t_d} + r_f E_{UV}
\label{OIRRate}
\end{equation}
where $t_d$ is the typical OIR diffusion time. Here we have ignored the
possibility that deposition results in the direct emission of OIR photons.
If optical photons are processed on a short enough time scale -- for
instance, if $t_d \lesssim t_e$ -- then we can set $\partial
E_{OIR}/\partial t\approx 0$ and solve for $E_{OIR}$:
\begin{equation}
E_{OIR} = {r_f \over 1/t_e + 1/t_d + r_f} E_{UV}.
\label{EOIREUV}
\end{equation}
Finally, by adding equations (\ref{UVRate}) and (\ref{OIRRate}) together, and
making use of equation (\ref{EOIREUV}), we obtain an equation for the total
energy density, $E_{tot}$:
\begin{equation}
{\partial E_{tot}\over\partial t} = \rho\dot{S}_{dep}(t) -
{E_{tot}\over t_e} - {E_{tot}\over \tilde{t}_d}
\label{TotRate}
\end{equation}
where the {\sl effective diffusion time}, $\tilde{t}_d$, is defined as
\begin{equation}
\tilde{t}_d \equiv {t_d + t_e + r_f t_d t_e \over r_f t_e}.
\label{tdeff}
\end{equation}
Written this way, it is easy to see that as $r_f$ -- the UV to OIR
fluorescence rate -- decreases, the effective diffusion time goes up. In
terms of the opacity, $r_f$ can be written in LTE as
\begin{equation}
r_f = c \rho
\left({\int_{\nu_{OIR}}^\infty \kappa_{\nu^\prime} B_{\nu^\prime}\,
d{\nu^\prime} \over \int_{\nu_{OIR}}^\infty B_{\nu^\prime}\,
d{\nu^\prime}}\right)
 \times
\left({\int_0^{\nu_{OIR}} \kappa_{\nu^\prime}
B_{\nu^\prime}\, d{\nu^\prime} \over \int_0^\infty \kappa_{\nu^\prime}
B_{\nu^\prime}\, d{\nu^\prime}}\right),
\end{equation}
where $B_\nu$ is the Planck function at the local gas temperature, and
$\nu_{OIR}$ is a frequency which approximately divides the UV from the
optical.  The first fraction is just the Planck mean opacity in the UV, and
the second fraction is the fraction of absorbed energy which is radiated at
optical wavelengths.  Since, for instance, Co~II has a larger opacity in
the optical than Co~III, increasing the ionization from once- to
twice-ionized will decrease the fraction of absorbed energy which is
re-radiated in the optical, making $r_f$ smaller, and therefore increasing
the effective diffusion time.

We can now understand the underlying physics of the WLR: as the amount of
\nifsx is increased, the interior temperature and ionization go up, which
decreases $r_f$ and increases $t_d$. Stated another way, brighter SNe~Ia
(\ie ones with more \nifsx) are hotter inside, and therefore have longer
diffusion times (\ie evolve more slowly through peak).

This cannot be the whole story, however. As long as most of the \nifsx is
unmixed and its mass fraction $X_{56}\approx 1$ in the \nifsx-dominated
layers, increasing the total amount of \nifsx in the star does not increase
the local heating rate per gram, and therefore would not, by itself, result
in a temperature increase.  Overlooked so far is the fact that, if the
additional \nifsx is added at higher velocity, then there will be an
increase in the effective optical depth.  That is, in stars with more
\nifsx (in a layered composition), the velocity of the outer edge of the
\nifsx is moving faster than is the outer edge of the \nifsx in a star with
less \nifsx. The higher the velocity of the outer edge of the \nifsx, the
more lines through which a photon must Doppler shift before escaping. This
represents an optical depth increase which would increase the escape time
and therefore make the gas hotter.  Putting iron-peak elements at higher
velocity is, by itself, not sufficient to increase $\tau_{eff}$ enough to
fully explain the observed correlation between light curve width and
luminosity. Rather, the slight increase in optical depth provided by
putting \nifsx at higher velocities is amplified by its effect on
temperature and ionization, and thus the escape time. It is difficult to
cleanly separate these two effects, but in the next section we present
results which strongly suggest that ionization is the dominant mechanism
determining the escape time.

It is worth stressing here that fluorescence is not a one-way process.
Energy can be transported to shorter wavelengths by ``combining'' photons.
A photon can be absorbed, raising an ion to an excited state. The ion can
then absorb another photon into this state, raising it to yet a
higher-energy state from which a UV photon can be emitted. From obvious
thermodynamic considerations (embodied in Rosseland's Theorem of Cycles)
this ``combining'' of photons must in general be less frequent than
``splitting'' of UV photons.  Near the peak in the lightcurve, however,
over much of the volume of the supernova these processes are nearly in
equilibrium; it is the {\em net} rate of transport of energy to longer
wavelengths which determines the escape time, not merely a one-way
fluorescent cascade. Indeed, \citet{Mazzali00} has suggested that
``combining'', in lower-optical depth regions above the photosphere, is a
significant contributor to the UV emission at maximum light.

Another avenue for time-dependence to enter the equations above is through
the energy deposition term $\dot{S}_{dep}(t)$. The deposition is determined
both by how much \nifsx is present and by what fraction of its decay energy
is deposited in the gas. Because the Compton opacity to MeV $\gamma$-rays
is mainly absorptive, the deposition fraction is roughly proportional to $1
- exp(-\tau_{\gamma})$, where $\tau_{\gamma}$ is the mean gamma-ray optical
depth and is in turn roughly proportional to the mean column depth to
\nifsx. More-rapid expansion will lead to a lower mean optical depth to
$\gamma$-rays at a given time, as will mixing radioactive material to
higher velocities.  Lower column depths will in turn lead to a more-rapid
decline in the deposition fraction.  If $\dot{S}_{dep}(t)$ declines more
rapidly, the lightcurve will peak earlier. An earlier peak will have more
\nifsx still present, and this will tend to offset the larger escape
fraction.  The lightcurve will thus evolve more rapidly, leading to a
narrower lightcurve for it luminosity.

We have thus identified three properties of the explosion which may
have a strong effect upon the WLR. The mass of \nifsx determines the
ionization and through it the mean escape time.  The expansion
velocity (or specific kinetic energy) determines the rate of decline
of the energy deposition. The radial distribution of \nifsx can have
an effect upon both the escape time of the thermalized radiation and
the time evolution of the deposition. In the following sections we
present a small set of synthetic lightcurves which illustrate these
effects and their relative importance.

\section{Computations}

A {\em complete} model for supernova lightcurves 
would begin with a model for the evolution of
the progenitor. This would be evolved to ignition, the subsequent explosive
burning followed to completion, and a transport calculation performed to
determine the lightcurve. Unfortunately, a fully predictive model is not
yet available from such a procedure. While there are schematic models for
all of these stages, it is not yet possible to simulate SN~Ia explosions
without introducing very simple parameterizations to stand in for very
complex physics we do not yet understand.

The evolution of the progenitor is undoubtedly the most uncertain part of
this process, yet there is the possibility that ``convergent evolution'' of
some sort, perhaps ignition near the Chandrasekhar mass in single degenerate
systems, reduces the sensitivity to initial conditions. On the other hand,
the initial conditions in the various double-degenerate scenarios might well
be the single most important piece of physics determining the subsequent
explosion. In the burning phase, the physics of turbulent combustion and
the possible spontaneous transition to detonation are probably the most
important and least tractable effects. While progress is being made on all
of these fronts \citep{ReineckeHN99,Iwamotoetal99}, it will be a long
time before truly predictive models of progenitor evolution and explosion
become available.

Our aim in this work is to explore the radiation transport effects which
influence the WLR. We have thus chosen to construct a very simple model for
the structure of the explosion. We start with model DD4 of
\citet{WoosleyW91}, an \Mch explosion which burned 1.26\Msun to Si or
heavier elements to yield $1.2\times 10^{51}$~ergs of kinetic energy,
producing 0.63 \Msun of \nifsx. We have modified the mass and
distribution of \nifsx in this model to explore the effects such
modifications have on the lightcurve. The density as a function of velocity
in the original model DD4 is preserved in all models. The structure of DD4
is presented by \citet{WoosleyW91} and by PE~II.

An initial grid of models with different \nifsx masses was obtained by
mapping between \nifsx and $^{28}$Si according to
\begin{eqnarray*}
X_{56}(M) & = & X^0_{56}(M) \exp\left(-\phi(M)\right) \\
X_{28}(M) & = & X^0_{28}(M) + X^0_{56}(M),
 \left(1 - \exp\left(-\phi(M)\right)\right)
\end{eqnarray*}
where
\begin{equation}
\phi(M) \equiv \max\left(\alpha(M-M_{cut}), 0\right).
\end{equation}

To {\sl increase} the \nifsx mass, all isotopes below $M=M_{cut}$ were
mapped to \nifsx:
\begin{equation}
X_{56}(M) =  1 - H(M-M_{cut}) + X^0_{56}(M) H(M-M_{cut}),
\end{equation}
where  $H(x)$ is the Heaviside function, and for $A\ne56$,
\begin{equation}
X_{A}(M)  =  X^0_{A}(M) H(M-M_{cut}).
\end{equation}

This simple scheme is actually well-motivated in a physical sense. The
total energy released from burning a C/O mixture to intermediate elements
in the Si group is very nearly the same as for burning to iron-peak
elements. The \nifsx mass thus need have little necessary connection to the
explosion kinetic energy. The most stringent constraints on the
distribution of elements in velocity come from peak-phase spectroscopy,
though this analysis is hampered by our inability to compute time-dependent
NLTE models.

As discussed in PE~I, the initial thermal energy in the explosion goes
entirely into kinetic energy via the second term on the r.h.s. of equation
(\ref{UVRate}) and hence has no effect upon the observable part of the
lightcurve. All the models presented here were started with the same
initial internal energy.

The lightcurves were computed as described in PE~II, using time-dependent,
multi-group radiation transport with an LTE equation of state and including
the elements He, C, O, Si, S, Ca, Fe, Co, and Ni in ionizations up to
Ni~XIII. 6000 energy groups were employed, providing moderate resolution;
doubling the number of groups does not significantly alter the results,
though halving it does leads to marked effects.  We are forced to
employ LTE level populations because of computer-time constraints -- a NLTE
calculation would take as long per time step as a whole lightcurve in LTE
(several days on a fast workstation). All current NLTE calculations
\citep{Hoflich95,NugentBBFH97,MazzaliL93} employ temperature structures
which are not consistent with the history of the ejecta's evolution and
therefore cannot be expected to give correct time-dependent spectra and
lightcurves, as discussed in PE~II. While these LTE calculations obviously
suffer as well, the ability to include time-dependent physics correctly is
evidently necessary to study the effects discussed above.


\begin{deluxetable}{lcccccccccccc}
\tiny
\tablecaption{MCLS Lightcurve Fits\label{widthtab}}
\tablecolumns{13}
\tablehead{
\colhead{model} & \colhead{M(\nifsx)} & \colhead{$M^{peak}_V$} & \colhead{$\Delta m_{15}$\tablenotemark{a}} & \colhead{$M^{peak}_V$\tablenotemark{b}}          & \colhead{$\Delta_{20}$\tablenotemark{c}} & \colhead{$M^{peak}_V$\tablenotemark{d}}       & \colhead{$\Delta_{40}$\tablenotemark{e}} & \colhead{$M^{peak}_V$\tablenotemark{f}}       & \colhead{$\Delta_{20}$\tablenotemark{g}} & \colhead{$\Delta_{40}$\tablenotemark{h}} \\
      & \colhead{(\Msun)}          & \colhead{(actual)}     &                  & \colhead{($\Delta m_{15}$)} &               & \colhead{($\Delta_{20}$)} &               & \colhead{($\Delta_{40}$)} &  \colhead{(BVR)}       &   \colhead{(BVR)}
}
\startdata
DD4/27 & 0.27 & -18.90 & 1.42 & -18.87 &  0.470 & -18.98 &  0.310 & -18.87 &  0.509 &  0.479 \\
DD4/38 & 0.38 & -19.17 & 1.47 & -19.15 &  0.348 & -19.10 &  0.241 & -19.21 &  0.460 &  0.427 \\
DD4/48 & 0.48 & -19.34 & 1.37 & -19.34 &  0.121 & -19.33 &  0.159 & -19.23 &  0.220 &  0.235 \\
DD4/55 & 0.55 & -19.42 & 1.24 & -19.43 & -0.054 & -19.51 &  0.123 & -19.47 &  0.160 &  0.157 \\
DD4    & 0.64 & -19.52 & 1.22 & -19.52 & -0.123 & -19.58 &  0.131 & -19.32 &  0.160 &  0.157 \\
DD4/90 & 0.90 & -19.78 & 0.87 & -19.81 & -0.443 & -19.91 & -0.542 & -20.02 & -0.704 & -0.710 \\
\tableline
M1     & 0.64 & -19.56 & 1.24 & -19.55 & -0.126 & -19.58 &  0.018 & -19.43 &  0.219 &  0.244 \\
NiFe   & 0.32 & -19.01 & 1.75 & -19.01 &  0.633 & -18.81 &  0.502 & -18.94 &  0.655 &  0.543 \\
DD3    & 0.93 & -19.90 & 0.87 & -19.90 &  0.083 & -19.36 & -0.161 & -19.62 & -0.698 & -0.525
\enddata
\tablenotetext{a}{$\Delta m_{15}$ measured from the V lightcurve.}
\tablenotetext{b}{Peak V magnitude as measured by the $\Delta m_{15}$ fit.}
\tablenotetext{c}{MCLS width measured from the V lightcurve from (-10,20) days.}
\tablenotetext{d}{Peak V magnitude as measured by the $\Delta_{20}$ fit.}
\tablenotetext{e}{MCLS width measured from the V lightcurve from (-10,40) days.}
\tablenotetext{f}{Peak V magnitude as measured by the $\Delta_{40}$ fit.}
\tablenotetext{g}{MCLS width measured from the B, V, and R lightcurves from (-10,20) days.}
\tablenotetext{h}{MCLS width measured from the B, V, and R lightcurves from (-10,40) days.}
\end{deluxetable}

The WLR is an {\em observed} relation between peak magnitude and a measure
of width. To obtain an accurate result in the face of measurement error and
the temporal sparsity of observed data, both the peak magnitude and the
width are measured by fitting observed data to templates determined from
well-observed supernovae \citep{Hamuyetal96a,RiessPK95}. It is important to
employ the same methods to measure the peak magnitude and width in our
models. We ``observe'' our synthetic lightcurves by employing the MCLS
(Multi-Color Lightcurve Shape) method of \citet{RiessPK95}, employing the
revised vectors used in \citet{Riessetal98} (kindly provided by
Dr. Riess). We have also determined the $\Delta\,m_{15}(B)$ decline-rate
parameter for our models using the $\chi^2$ template-fitting technique
described by \citet{Hamuyetal96d}. The results of fitting from both of
these techniques are shown in table \ref{widthtab}.

We note that the observational difficulty of obtaining early lightcurve
data has hindered the construction of templates for times earlier than ten
days before peak. Thus, by construction, template-fitting techniques are
most sensitive to post-maximum decline rates; their ability to discriminate
between different pre-maximum lightcurve shapes is seriously hampered.

Observationally, WLR relations with similar scatter exist in the B, V, R,
and I bands -- color evolution during the first 30 days is well correlated
with luminosity and width. At later times, all SNe~Ia show a remarkable
homogeneity in color. In the absence of reddening, determining the width of
the lightcurve in a single band is sufficient to uniquely determine a
supernova's peak magnitude.  The importance of fitting a supernova's color
evolution is primarily in determining extinction. As our models do not
suffer from this observational uncertainty, it is sufficient to fit to a
single band.

We remind the reader as well that individual supernovae show significant
departures from the best-fit templates (\cf \citealp{Hamuyetal96c}); SNe~Ia
do {\em not} form a strictly one-parameter family. While the width of
lightcurves measured by the templates is the largest variation from object
to object, there are other features in the lightcurve which are not
correlated with the decline rate. A good example of this is the pair
SN1991T and SN1992bc which have very similar decline rates over the first
15 days past peak but very different rates over the subsequent 45 days
\citep{Hamuyetal96d}.

The details of the atomic physics and NLTE effects determine the color
evolution of the supernova. We have assumed in this study that such effects
are less important to the bolometric luminosity, merely reapportioning flux
between the B, V, and R bands which comprise most of the luminosity
\citep{Suntzeff95}.  We have thus chosen to take fits to the WLR in
the V band only as most representative of the overall relation between
width and luminosity. Below we present fits to B, V, and R simultaneously
as well, but as an accurate determination of the color evolution depends
more upon specific details of the atomic models and NLTE effect, we regard
these fits as less reliable indicators of the underlying physics than the
trends in a single band.

As the ejecta evolve, the effects of trapping and fluorescence become less
important to the bolometric lightcurve. The {\em colors} at these later
times, however, are very strongly influenced by NLTE effects. LTE
calculations cannot be expected to provide an adequate representation of
the lightcurve physics as the supernova becomes a nebula. Thus, we fit only
to the peak phase of the lightcurve and estimate the sensitivity of fitting
to various time intervals by fitting from 10 days before peak (the earliest
time in the templates) until 20 and 40 days past peak.

\section{Results}

\begin{figure}[!t]
\epsscale{0.8}
\plotone{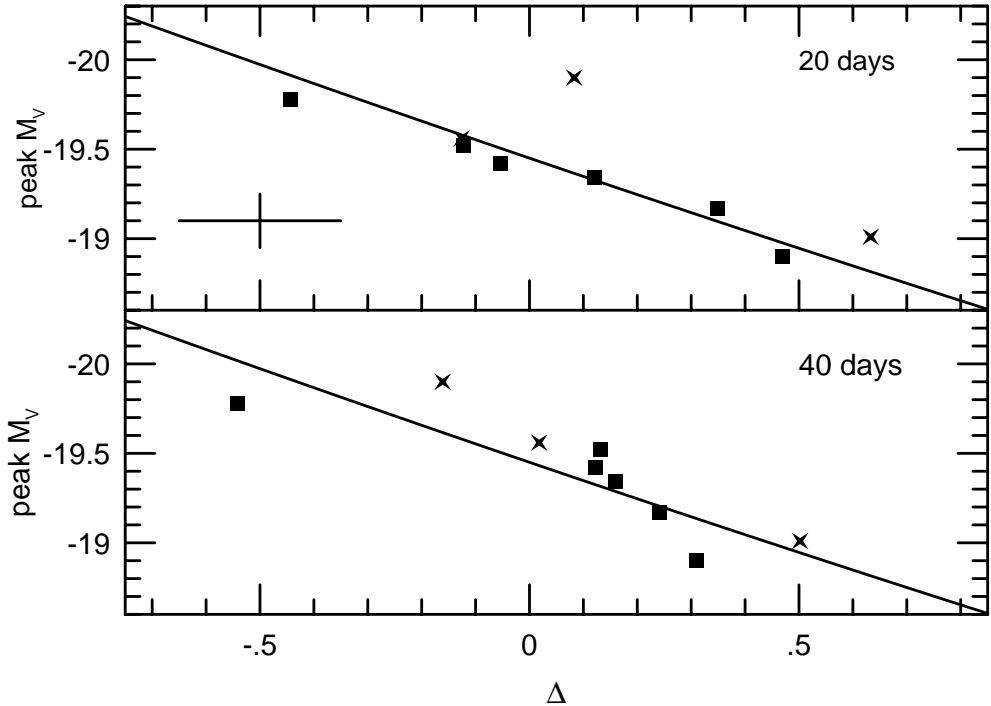}
\caption{Peak V magnitudes {\it vs} the MCLS $\Delta$ parameter (a measure
of lightcurve width) fit to V lightcurves over the interval (-10,20) days
(top panel) and (-10,40) days (bottom panel) from peak.  The squares
correspond to calculations made by varying the \nifsx mass in model
DD4. The stars show the three variations from these models as discussed in the
text. The diagonal curve in each panel is the observed quadratic relation
between $M_V^{peak}$ and $\Delta$ using the Cepheid calibration of
SNe~Ia. The approximate magnitude of the observed dispersion about this
relation, corresponding to $\pm0.15$ mag and $\pm0.125$ in $\Delta$, is
indicated in the top panel}
\label{magvsdelta}
\end{figure}

Figure (\ref{magvsdelta}) compares the MCLS $\Delta$ parameter (a measure
of width) to the peak V absolute magnitude determined from the models
described above. The MCLS method parameterizes the shape of a supernova's
lightcurve as
\begin{equation}
M(t) = A(t) + \Delta \times L(t) + \Delta^2 \times Q(t)
\end{equation}
where $A(t)$ is, roughly, the mean lightcurve shape of a sample of 27
well-observed supernovae \citep{Riessetal98} and $L(t)$ and $Q(t)$ are
functions describing a quadratic fit to the departure of individual
supernovae from this mean shape as parameterized by $\Delta$ which is thus
a measure of lightcurve width. The quadratic relation between $\Delta$ and
$V_{peak}$ is indicated by the diagonal curves in the figure. The absolute
luminosity of the relation is calibrated by the Cepheid distance scale of
\citep{Gibsonetal00} corresponding to a Hubble constant of
68~km~s$^{-1}$~Mpc$^{-1}$. The square symbols correspond to the models with
0.27, 0.38, 0.48, 0.55, 0.63, and 0.90 M\sun of \nifsx in the structure of
model DD4. The star symbols represent other models to be discussed below.
The top panel shows fits over the interval (-10,20) days relative to peak
and the lower panel over the interval (-10,40). The quality of the fit to
the mean WLR relation is somewhat better for the scaled-\nifsx models for
the shorter interval and is probably a consequence of the increasingly
inappropriate assumption of LTE at later times.

In both cases, the fits show a dispersion comparable to or smaller than the
observations ($\sim0.15$~mag) and (remarkably and probably fortuitously)
virtually zero departure from the absolute calibration.  It is clear from
this figure that the V WLR is fit remarkably well by the simple
prescription of changing only the \nifsx mass.

\begin{figure}[!t]
\plotone{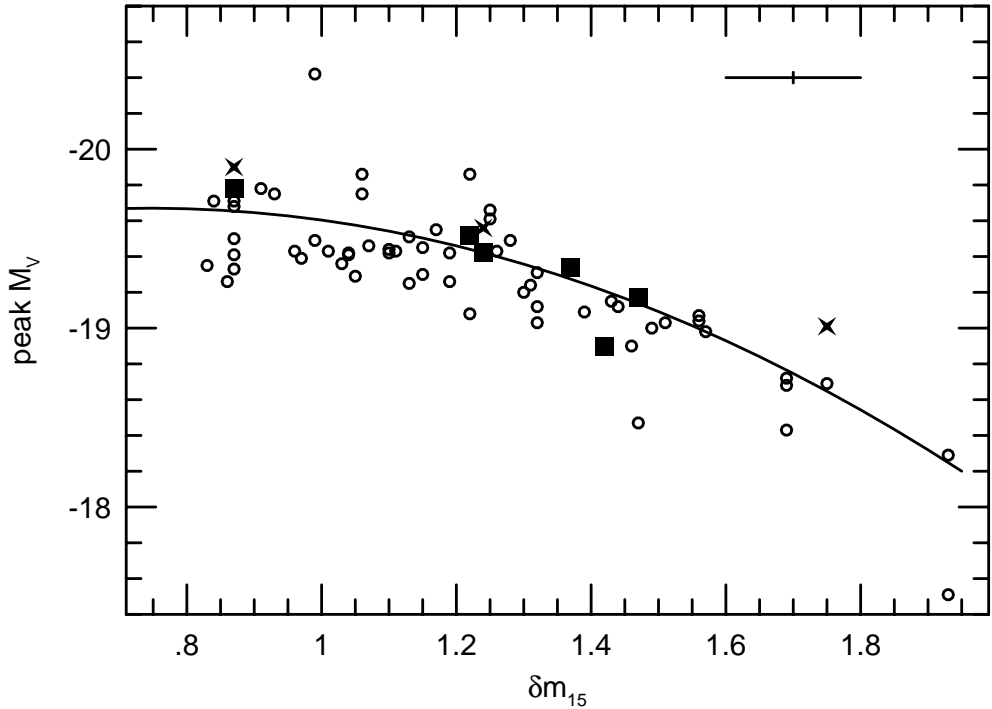}
\caption{Peak V magnitudes {\it vs} the $\Delta\,m_{15}(B)$ parameter
(another measure of lightcurve width) measured from V lightcurves and on
the same luminosity calibration as the previous figure.  Open symbols are
data for 59 supernovae from \protect\citet{Phillipsetal98}, on the same
calibration as in the previous figure. The stars show the three variations
from the scaled-DD4 models as discussed in the text.}
\label{mariofig}
\end{figure}

Figure (\ref{mariofig}) shows the same lightcurves fit by the
template-fitting ($\Delta\,m_{15}(B)$) method of \citet{Hamuyetal96a} on the
interval (-5,26) days. In this method, the models are fit to each of six
template lightcurves spanning the range of well-observed supernovae. The
templates are characterized by their $\Delta\,m_{15}(B)$ parameters (roughly
the decline in B magnitude over 15 days from peak), while the actual epochs
fit are determined by the whole peak phase of the lightcurves. The best-fit
$\Delta\,m_{15}(B)$ is then determined from the minimum of a quadratic fit to
the reduced $\chi^2$ as a function of the parameter. Rather than
extrapolating, models broader than the broadest template are assigned the
minimum value of $\Delta\,m_{15}(B)$, accounting for the slight pile-up of
points at the bright end of the relation.

The filled symbols are the same as those of the previous figure, and the
open symbols are fits to 59 well-observed supernovae from
\citet{Phillipsetal98}. The curve is the mean $\Delta\,m_{15}(B)$-$M_V$
relation, on the same calibration, from \citep{Phillipsetal98}. Once again,
the models fit the relation well within the scatter of observed
supernovae. The significant difference in curvature between the mean
$\Delta\,m_{15}(B)$-luminosity and $\Delta$-luminosity relations is notable
and appears to be the result of the differing ways in which the lightcurve
shapes are parameterized; both techniques achieve similar reductions in the
scatter about the Hubble relation.

\begin{figure}[!t]
\plotone{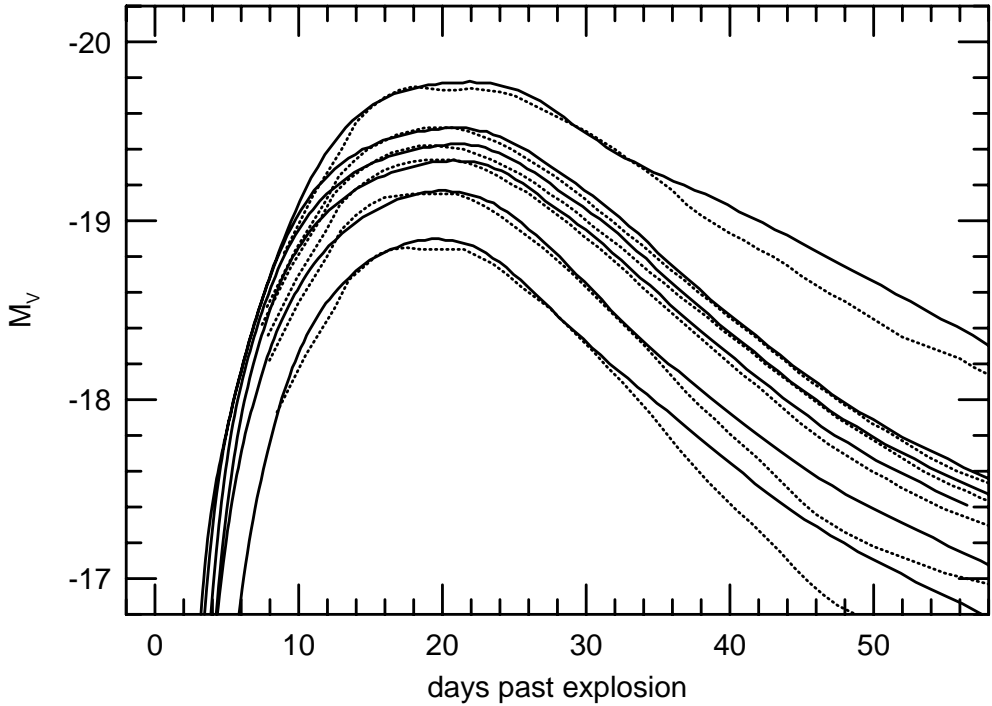}
\caption{The six V scaled-DD4 lightcurves. The solid lines are the
synthetic lightcurves; the dotted lines are the best-fit MCLS template
lightcurves fit over the interval (-11,20) days with respect to peak.}
\label{vlcs}
\end{figure}

Figure (\ref{vlcs}) shows the V lightcurve family from the six scaled DD4
models compared with the best-fitting MCLS templates. The shapes of the
lightcurves depart less than 0.05 magnitude from the templates throughout
the first 45 days (about 25 days past V peak) with the exception of the
0.90 M\sun model, where the inflection at 32 days is not seen in the
templates. This error is well within the dispersion of data from individual
supernovae about the template lightcurves (\cf \citealp{Riessetal98}).  At
later times, the LTE approximation becomes increasingly poor and may
account for the too-shallow decline rates, though real supernovae show
similar departures from the templates at this time.

\begin{figure}[!t]
\plotone{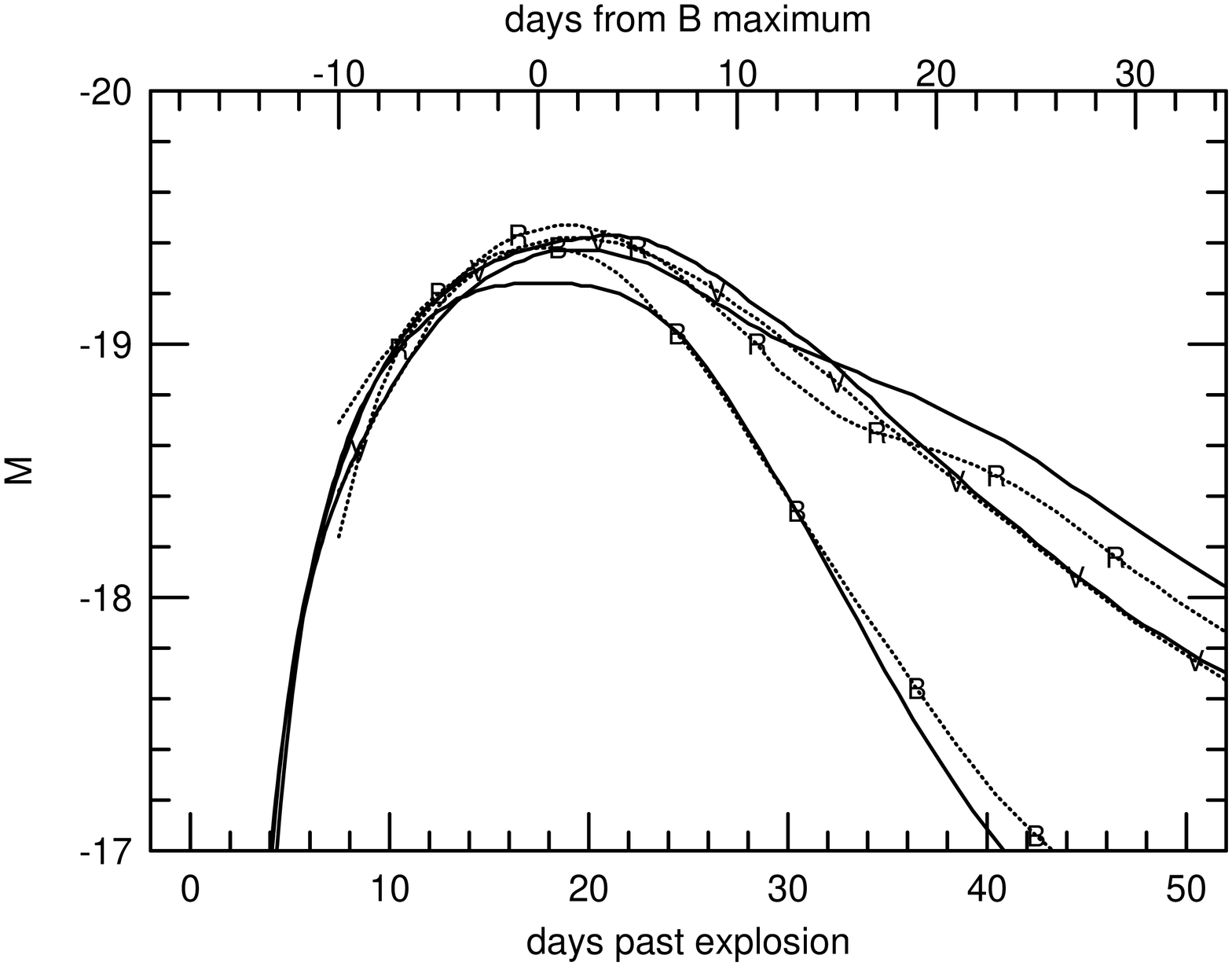}
\caption{B, V, and R lightcurves of model DD4/55 and the best-fit MCLS
templates on the interval (-10,20) days relative to peak.}
\label{plot55}
\end{figure}

\begin{figure}[!t]
\plotone{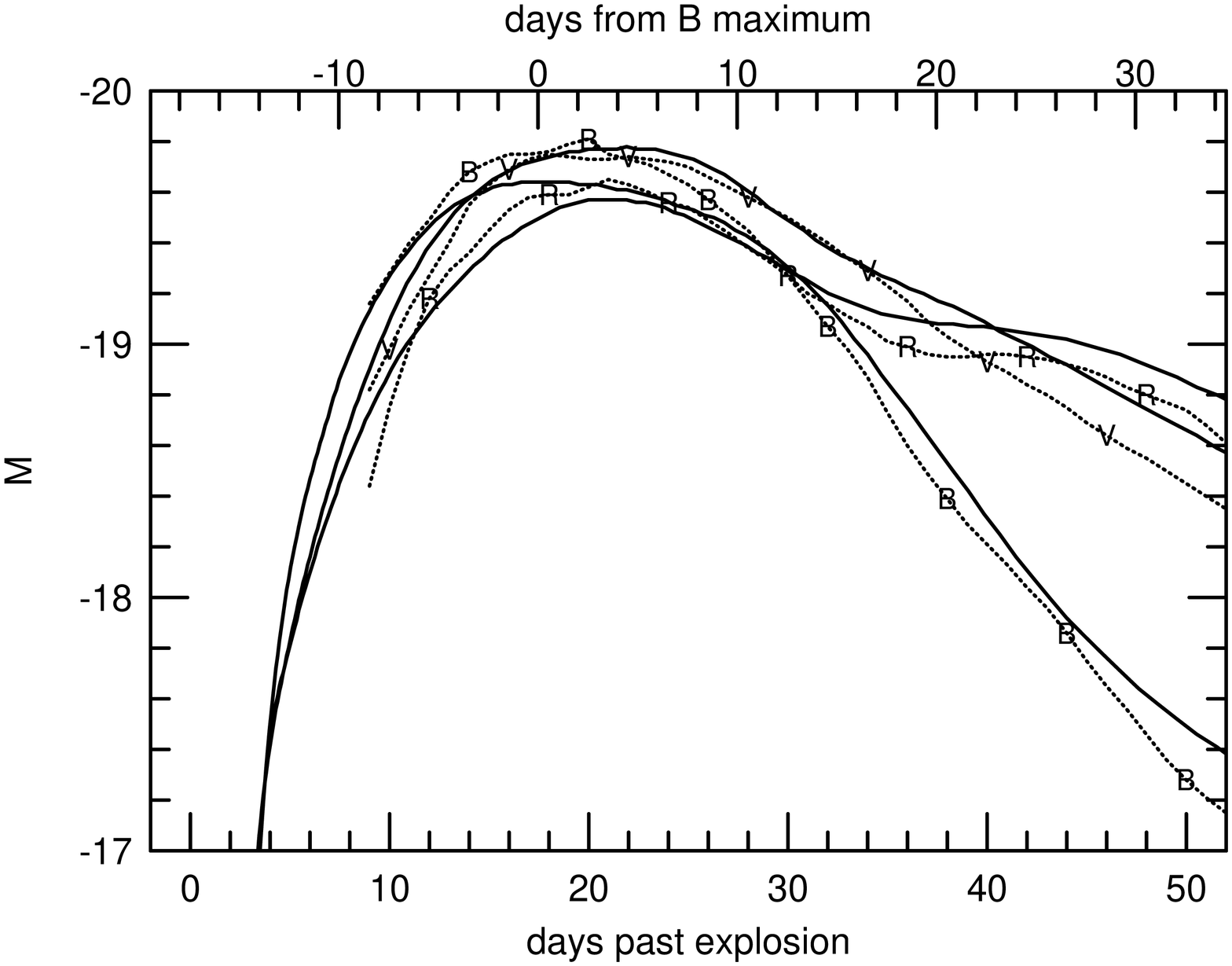}
\caption{B, V, and R lightcurves of model DD4/90 and the best-fit MCLS
templates on the interval (-10,20) days relative to peak.}
\label{plot90}
\end{figure}

Figures (\ref{plot55}) and (\ref{plot90}) show the B, V, and R lightcurves
of the 0.55 M\sun and 0.90 M\sun models, respectively.  The DD4/55 model
displays a generic problem with most of the models; while the B lightcurve
is the correct width, near peak it is too flat compared to the templates.
The departure from the templates before peak is less disturbing, and is
probably more a problem with the templates than the models, as there is
comparatively little data so far before maximum. The IR secondary maximum
is clearly seen in the R-band curves.

We now turn to variations on these scaled-DD4 models to investigate the
sensitivity of the WLR to other changes in the explosion.  The models
presented so far have nearly pure \nifsx from the center out to a mass
coordinate nearly equal to the Ni mass -- radioactive material is
concentrated as much as possible toward the center.  This is a natural
result of the explosion physics. White dwarfs are densest at the
center. When matter burns explosively, the higher the density, the larger
the ratio of the hydrodynamic timescale to the the burning timescale, and
the more the distribution of burning products is skewed toward heavier
elements. In the absence of significant hydrodynamic mixing after material
has cooled below the burning temperature, a natural consequence of
explosive burning is that the gradient of atomic weight will point inward,
with the heaviest burning product, \nifsx, concentrated toward the core.

To explore the effect hydrodynamic mixing might have on the
lightcurve, we have altered the distribution of \nifsx in model DD4
without changing its mass. A significant mass fraction of \nifsx
extends approximately to 0.85\Msun in the original model (primarily as
a consequence of producing 0.16\Msun of non-radioactive iron-group
material).  The starred points at $M_V = 19.56$ in Figures
(\ref{magvsdelta}) and (\ref{mariofig}) show the result of
homogenizing the composition of this model over the mass range 0.7 to
1.1 M\sun in model M1.  This decreases the peak bolometric magnitude
by 0.1 mag and leads to significantly narrower B and brighter I
lightcurves.  The resulting V and R magnitudes are virtually
identical, 0.04 and 0.02 mag brighter, respectively.  The \nifsx
nearer the surface cooled more rapidly. The lower ionization gas
converts energy from shorter- to longer-wavelength photons more
efficiently, circumventing the effect of the higher expansion optical
depth from \nifsx at higher velocities. Altering the distribution of
\nifsx would appear to affect the color of the emergent flux more than
its overall escape properties. An important conclusion is that, within
the limits of fitting 0.63\Msun of \nifsx in an \Mch white dwarf, the
WLR is insensitive to the distribution of radioactivity in the
explosion.

The scaled-DD4 models traded off production of \nifsx with
intermediate mass elements. This is the only possibility afforded if
the supernovae are to satisfy nucleosynthetic constraints. For
example, the production of large quantities of non-radioactive
iron-group elements in the majority of supernov\ae\ is firmly ruled
out.  Nonetheless, in model NiFe we have converted, in place, half of
the mass of \nifsx to \fefsx, thus changing the heating rate per gram
of the iron-rich material without changing its distribution, cooling,
or $\gamma$-ray escape properties. The lower temperature in the
\nifsx-rich material leads to lower ionization and hence to more
efficient escape. In spite of having 0.32 instead of 0.27\Msun of
\nifsx, the model has a much narrower lightcurve than model DD4/27, as
shown by the stars near $M_V = -19.01$ in the Figures~1 and 2.  In model M1
\nifsx was mixed outward with overlying Si-group material. However the
amount of material involved in the mixing was a small fraction of the
total mass of \nifsx. Consequently the effectiveness of mixing in
lowering the heating rate per gram and narrowing the lightcurve
was small.

Interchanging Fe- with Si-group elements makes relatively little difference
in the amount of energy released by burning a C/O mixture; at most
$6.2\times 10^{17}$\ergg is released burning to Si \& S {\it versus}
$7.9\times10^{17}$\ergg for burning to NSE. It is thus quite possible to
imagine a series of explosions in which the Si/Fe ratio varies more
strongly than the specific kinetic energy. On the other hand, in most
current models a greater production of \nifsx is accompanied by a greater
total amount of burning, leading to higher velocities. Model DD3 of
\citet{WoosleyW91} produced 0.96 \Msun of \nifsx. It burned 10\% more mass
to the Si-group or above in total and has 10\% greater kinetic energy than
model DD4.  In spite of its high \nifsx mass, the lightcurve from this
model is considerably narrower than that of model DD4/90 and is shown by
the stars in Figures (\ref{magvsdelta}) and (\ref{mariofig}) at
$M_V=-19.90$.  The higher kinetic energy in this explosion resulted in a
lower column depth, allowing decay $\gamma$-rays to escape more readily
after peak.  By 20 days, $\sim6$\% of the decay energy in model DD4/90
escaped directly as $\gamma$-rays, increasing to 35\% by day 40. In model
DD3, the escape fraction was more than double at 20 days, at $\sim$14\%,
rising to 46\% by day 40.  A shortened time to $\gamma$-ray transparency
leads to a shorter rise-time (by 3.5 days), a narrower peak, and a greater
decline rate in the lightcurve. The increased $\gamma$-ray escape was,
however, more than compensated for by the shorter rise-time, and the peak
magnitude is even brighter than the difference one might predict based
solely on consideration of the difference in total \nifsx mass.  It is
interesting to note that the $\Delta\,m_{15}(B)$ fitting procedure did not
pick up this variation in lightcurve shape as it weights more heavily the
decline rate after peak than the MCLS templates; the two models had very
similar post-maximum declines.

\section{Conclusions}

While far from a thorough investigation of all the variations one can
imagine on explosions of \Mch white dwarfs, the results presented above
are suggestive. They show that it is possible to create a series of models
which exhibit, to the best of our current calculation ability yet well
within observational errors, the observed WLR.

It seems clear from the arguments of \S2 and the results of
these simple experiments that the \nifsx yield and the $\gamma$-ray escape
fraction play the determining roles in the relation between the
lightcurve's width and peak luminosity. The models show a remarkable
insensitivity to the distribution of radioactivity and to the composition
of non-radioactive material, within limits set by achieving the appropriate
explosion energies and acceptable nucleosynthesis.

The mass of \nifsx sets the scale of the peak luminosity and the heating
rate of the gas. The resulting temperature determines the ionization and
hence the cooling rate of the gas. More \nifsx leads to more heating,
higher temperatures, less efficient cooling, and hence broader and brighter
lightcurves.

The efficiency of $\gamma$-ray escape sets the evolution rate of the energy
deposition.  Models with higher overall velocities and/or with \nifsx
distributions extending to higher velocities exhibit larger $\gamma$-ray
escape fractions at a given time. This increases the evolution rate of the
lightcurve, with greater escape leading to narrower lightcurves.  Models
with higher \nifsx yields tend to have their radioactivity distributed over
a broader range in velocity, leading to larger escape fractions.  This
greater escape fraction acts to oppose the tendency for broader lightcurves
resulting from higher temperatures, decreasing the slope of the WLR.

We have reproduced the observed relation in V by varying {\em only} the
\nifsx yield. One solution to reproducing the observed WLR is thus to
produce an evolution scenario which leads to supernov\ae\ with similar
explosion energies yet differing \nifsx yields. Because higher \nifsx
yields and higher velocities tend to produce opposing effects on the
lightcurve, this solution is almost certainly not unique. A weak dependence
of energy on \nifsx yield will likely lead to acceptable lightcurves as
well.

While to date the models which best fit lightcurves and spectra are all at
the Chandrasekhar mass (with the possible exception of models for
SN1991bg), the masses of all SN~Ia explosions may not be identical. Varying
the explosion mass primarily affects the $\gamma$-ray optical depth. As
long as a given \nifsx mass is accompanied by a column depth similar to
those exhibited by the \Mch models considered here, a variable-mass
explosion scenario would not appreciably alter our conclusions. For
sub-\Mch model such as those of \citep{LivneA95,WoosleyW94}, the \nifsx
produced by detonation in surface layers does not contribute appreciably to
the optical lightcurve due to the low $\gamma$-ray optical depth. In such
models, only the \nifsx produced by carbon detonation in the core C/O dwarf
would affect the WLR.

The exact values of the slope and calibration of the WLR obtained in this
work must be, to some extent, fortuitous. EP~II showed that the conditions
experienced by matter in SNe~Ia are not consistent in detail with the
assumption of LTE. The shape and colors of the lightcurve depend
sensitively upon the average branching ratio from UV to OIR photons through
a broad range in energies. While many strong lines are unlikely to be
missing from our atomic models, all such models are seriously incomplete
and the statistical properties of the line distributions and branching
ratios are unlikely to be correct in detail. The basic physics, however,
will not change significantly, and thus we have some confidence in the
general trends the models indicate.

By fitting our models using the same procedures used to calibrate observed
SNe~Ia, we have demonstrated that the calibration is insensitive to fairly
large changes in the underlying explosions. This suggests that the small
differences in initial conditions which might arise from evolutionary
effects between the local universe and that at $z\sim1$ are unlikely to
appreciably affect the supernova cosmology results. For example, mixing in
model M1 leads to much less than 0.1 magnitude difference from the
corresponding unmixed model. Even the unphysically large change in
abundances in model NiFe, strongly ruled out by chemical evolution
considerations, leads to less than 0.2 magnitude departures from the MCLS
and barely 0.3 magnitude departures from the $\Delta\,m_{15}(B)$ relations;
allowable changes in relative iron-peak abundances would lead to
insignificant effects.

Altering the specific kinetic energy, as in Model DD3, is the only way we
see to achieve significant departures from the WLR. Increasing the kinetic
energy by 10\% was seen to lead to a much narrower lightcurve (at least
when fit by the MCLS templates), primarily by more than doubling the
$\gamma$-ray escape from 6\% at peak in DD4/90 to 14\% in DD3.  This change
is in the wrong direction, however.  To account for the 0.3 magnitude {\em
deficit} in brightness seen at large redshifts, a shift to lower explosion
energy would be required.  Halving the $\gamma$-ray escape from 6\% at peak
to 3\% is a smaller fractional change in deposition and would lead to a
smaller change in width. A future paper will explore these effects in
greater detail.

\acknowledgments
We wish to thank M. Hamuy for many useful discussions and
for performing the $\Delta\,m_{15}(B)$ fits to the lightcurves, and A. Riess
for supplying his lightcurve templates in advance of publication.
This work has been supported by the National Science Foundation
(CAREER grant AST9501634), by the National Aeronautics and Space
Administration (grant NAG~5-2798). This work was also performed under
the auspices of the U.S. Department of Energy by University of
California Lawrence Livermore National Laboratory under contract
No. W-7405-Eng-48.  Philip Pinto gratefully acknowledges support from
the Research Corporation though a Cottrell Scholarship.


\begin{thebibliography}{}

\bibitem[\protect\citeauthoryear{{Arnett}}{{Arnett}}{1982}]{Arnett82}
{Arnett}, W.~D. 1982, \apj, 253, 785

\bibitem[\protect\citeauthoryear{{Castor}, {Abbott}, \& {Klein}}{{Castor}
  et~al.}{1975}]{CastorAK75}
{Castor}, J., {Abbott}, D.,  \& {Klein}, R. 1975, \apj, 195, 157

\bibitem[\protect\citeauthoryear{{Eastman} \& {Pinto}}{{Eastman} \&
  {Pinto}}{1993}]{EastmanP93}
{Eastman}, R.~G.,  \& {Pinto}, P.~A. 1993, \apj, 412, 731

\bibitem[\protect\citeauthoryear{{Garnavich} et~al.}{{Garnavich}
  et~al.}{1998}]{Garnavichetal98}
{Garnavich}, P.~M., et~al. 1998, \apj, 509, 74

\bibitem[\protect\citeauthoryear{{Gibson} et~al.}{{Gibson}
  et~al.}{2000}]{Gibsonetal00}
{Gibson}, B.~K., et~al. 2000, \apj, 529, 723

\bibitem[\protect\citeauthoryear{{Hamuy} et~al.}{{Hamuy}
  et~al.}{1996a}]{Hamuyetal96c}
{Hamuy}, M., et~al. 1996a, \aj, 112, 2408

\bibitem[\protect\citeauthoryear{{Hamuy} et~al.}{{Hamuy}
  et~al.}{1996b}]{Hamuyetal96a}
{Hamuy}, M., {Phillips}, M.~M., {Suntzeff}, N.~B., {Schommer}, R.~A., {Maza},
  J.,  \& {Aviles}, R. 1996b, \aj, 112, 2391

\bibitem[\protect\citeauthoryear{{Hamuy} et~al.}{{Hamuy}
  et~al.}{1996c}]{Hamuyetal96d}
{Hamuy}, M., {Phillips}, M.~M., {Suntzeff}, N.~B., {Schommer}, R.~A., {Maza},
  J.,  \& {Aviles}, R. 1996c, \aj, 112, 2398

\bibitem[\protect\citeauthoryear{{Hamuy} et~al.}{{Hamuy}
  et~al.}{1996d}]{Hamuyetal96b}
{Hamuy}, M., {Phillips}, M.~M., {Suntzeff}, N.~B., {Schommer}, R.~A., {Maza},
  J., {Smith}, R.~C., {Lira}, P.,  \& {Aviles}, R. 1996d, \aj, 112, 2438

\bibitem[\protect\citeauthoryear{{H\"oflich}}{{H\"oflich}}{1995}]{Hoflich95}
{H\"oflich}, P. 1995, \apj, 443, 89

\bibitem[\protect\citeauthoryear{{H\"oflich} et~al.}{{H\"oflich}
  et~al.}{1996}]{Hoflichetal96}
{H\"oflich}, P., {Khokhlov}, A.~M., {Wheeler}, J.~C., {Phillips}, M.~M.,
  {Suntzeff}, N.~B.,  \& {Hamuy}, M. 1996, \apjl, 472, L81

\bibitem[\protect\citeauthoryear{{Iwamoto} et~al.}{{Iwamoto}
  et~al.}{1999}]{Iwamotoetal99}
{Iwamoto}, K., {Brachwitz}, F., {Nomoto}, K.~I., {Kishimoto}, N., {Umeda}, H.,
  {Hix}, W.~R.,  \& {Thielemann}, F.~K. 1999, \apjs, 125, 439

\bibitem[\protect\citeauthoryear{{Karp} et~al.}{{Karp}
  et~al.}{1977}]{KarpLCS77}
{Karp}, A.~H., {Lasher}, G., {Chan}, K.~L.,  \& {Salpeter}, E.~E. 1977, \apj,
  214, 161

\bibitem[\protect\citeauthoryear{{Kowal}}{{Kowal}}{1968}]{Kowal68}
{Kowal}, C.~T. 1968, \aj, 73, 1021

\bibitem[\protect\citeauthoryear{{Livne} \& {Arnett}}{{Livne} \&
  {Arnett}}{1995}]{LivneA95}
{Livne}, E.,  \& {Arnett}, D. 1995, \apj, 452, 62


\bibitem[\protect\citeauthoryear{{Mazzali}, P.A.}
{{Mazzali}}{2000}]{Mazzali00}
{Mazzali}, P.A., 2000, preprint.


\bibitem[\protect\citeauthoryear{{Mazzali} \& {Lucy}}{{Mazzali} \&
  {Lucy}}{1993}]{MazzaliL93}
{Mazzali}, P.~A.,  \& {Lucy}, L.~B. 1993, \aap, 279, 447

\bibitem[\protect\citeauthoryear{{Nugent} et~al.}{{Nugent}
  et~al.}{1997}]{NugentBBFH97}
{Nugent}, P., {Baron}, E., {Branch}, D., {Fisher}, A.,  \& {Hauschildt}, P.~H.
  1997, \apj, 485, 812

\bibitem[\protect\citeauthoryear{{Perlmutter} et~al.}{{Perlmutter}
  et~al.}{1999}]{Perlmutteretal99a}
{Perlmutter}, S., et~al. 1999, \apj, 517, 565

\bibitem[\protect\citeauthoryear{{Perlmutter}, {Turner}, \&
  {White}}{{Perlmutter} et~al.}{1999}]{PerlmutterTW99}
{Perlmutter}, S., {Turner}, M.~S.,  \& {White}, M. 1999, Physical Review
  Letters, 83, 670

\bibitem[\protect\citeauthoryear{{Perlmutter}}{{Perlmutter}}{1997}]{Perlmutter%
etal97}
{Perlmutter}, S. e.~a. 1997, \apj, 483, 565

\bibitem[\protect\citeauthoryear{{Phillips}}{{Phillips}}{1993}]{Phillips93}
{Phillips}, M.~M. 1993, \baas, 182, 2907

\bibitem[\protect\citeauthoryear{{Phillips} et~al.}{{Phillips}
  et~al.}{1999}]{Phillipsetal98}
{Phillips}, M.~M., {Lira}, P., {Suntzeff}, N.~B., {Schommer}, R.~A., {Hamuy},
  M.,  \& {Maza}, J. 1999, \aj, 118, 1766

\bibitem[\protect\citeauthoryear{{Pinto} \& {Eastman}}{{Pinto} \&
  {Eastman}}{2000a}]{PintoE00b}
{Pinto}, P.,  \& {Eastman}, R. 2000a, \apj, 001, in press

\bibitem[\protect\citeauthoryear{{Pinto} \& {Eastman}}{{Pinto} \&
  {Eastman}}{2000b}]{PintoE00a}
{Pinto}, P.~A.,  \& {Eastman}, R.~G. 2000b, \apj, 000, in press

\bibitem[\protect\citeauthoryear{{Reinecke}, {Hillebrandt}, \&
  {Niemeyer}}{{Reinecke} et~al.}{1999}]{ReineckeHN99}
{Reinecke}, M., {Hillebrandt}, W.,  \& {Niemeyer}, J.~C. 1999, \aap, 347, 739

\bibitem[\protect\citeauthoryear{{Riess} et~al.}{{Riess}
  et~al.}{1998}]{Riessetal98}
{Riess}, A., et~al. 1998, \aj, 116, 1009

\bibitem[\protect\citeauthoryear{{Riess}, {Press}, \& {Kirshner}}{{Riess}
  et~al.}{1995}]{RiessPK95}
{Riess}, A.~G., {Press}, W.~H.,  \& {Kirshner}, R.~P. 1995, \apjl, 438, L17

\bibitem[\protect\citeauthoryear{{Schmidt} et~al.}{{Schmidt}
  et~al.}{1998}]{Schmidtetal98}
{Schmidt}, B.~P., et~al. 1998, \apj, 507, 46

\bibitem[\protect\citeauthoryear{{Suntzeff}}{{Suntzeff}}{1995}]{Suntzeff95}
{Suntzeff}, N.~B. 1995, in IAU colloquium 145: Supernovae and Supernova
  Remnants, ed. R.~{McCray} (Cambridge: Cambridge University Press)

\bibitem[\protect\citeauthoryear{{Wagoner}, {Perez}, \& {Vasu}}{{Wagoner}
  et~al.}{1991}]{WagonerPV91}
{Wagoner}, R.~V., {Perez}, C.~A.,  \& {Vasu}, M. 1991, \apj, 377, 639

\bibitem[\protect\citeauthoryear{{Woosley} \& {Weaver}}{{Woosley} \&
  {Weaver}}{1991}]{WoosleyW91}
{Woosley}, S.~E.,  \& {Weaver}, T.~A. 1991, in Les Houches, Session LIV, ed.
  J.~{Audouze}, S.~{Bludman}, R.~{Mochkovitch}, \& J.~{Zinn-Justin} (Elsevier
  Science Publishers)

\bibitem[\protect\citeauthoryear{{Woosley} \& {Weaver}}{{Woosley} \&
  {Weaver}}{1994}]{WoosleyW94}
{Woosley}, S.~E.,  \& {Weaver}, T.~A. 1994, \apj, 423, 371
\end{thebibliography}

\end{document}